\documentclass{article}

\usepackage{arxiv}

\usepackage[utf8]{inputenc} 
\usepackage[T1]{fontenc}    
\usepackage{hyperref}       
\usepackage{url}            
\usepackage{booktabs}       
\usepackage{amsfonts}       
\usepackage{nicefrac}       
\usepackage{microtype}      
\usepackage{lipsum}		
\usepackage{graphicx}
\usepackage{natbib}
\usepackage{doi}

\title{On the luminosity-angular distances relation for an expanding universe}

\author{ \href{https://orcid.org/0000-0001-8318-6813}{\includegraphics[scale=0.06]{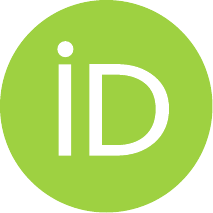}\hspace{1mm}Juan De Vicente}\\
	Centro de Investigaciones Energ\'eticas, Medioambientales y Tecnol\'ogicas (CIEMAT),\\Avda. Complutense 40, E-28040, Madrid, Spain\\
              \texttt{juan.vicente@ciemat.es} }

\renewcommand{\shorttitle}{On the luminosity-angular distances relation for an expanding universe}

\hypersetup{
pdftitle={Expansion Lensing I},
pdfsubject={astrophysics, cosmology},
pdfauthor={Juan De Vicente},
pdfkeywords={cosmological parameters, distances and redshift},
}

\begin{document}
\maketitle

\begin{abstract}
The luminosity-angular distance relation for an expanding universe is given by $D_L=D_A(1+z)^2$. This relation is commonly proven by geometrical considerations from the source point of view assuming an expanding universe. In this note, the same cosmological relation is derived from the observer point of view for the assumed expanding universe.
\end{abstract}


\section{Introduction}

    Observational signs of a possible expansion of the universe were found by ~\citet{hubble1929relation}; the correlation between redshifts and distances for extragalactic sources was interpreted as the evidence of an expanding universe. Different cosmological tests were proposed to probe whether the Universe is expanding or remains static. Tolman (1930, 1934) predicted that in an expanding universe, the surface brightness of a receding source will drop with redshift $z$ as $\sim(1+z)^{-4}$. Consequently to Tolman's prediction, the equation $D_L=D_A(1+z)^2$ was established between \textit{luminosity distance} $D_L$ and \textit{angular diameter distance} $D_A$. Another test for universe expansion is the time dilation of Type Ia supernovae light curves that was suggested by ~\citet{wilson1939} and assumed nonsense for a static universe. The results obtained by ~\citet{leibundgut1996time} and ~\citet{goldhaber2001timescale} on this test supports such time dilation, but not necessarily an expanding universe since it also may occur in a static universe.
    
     Deduction of the standard luminosity-angular distance relation from the source point of view assuming an expanding universe is common and can be found elsewhere. In this note, we derive the same relation from the observer point of view for an assumed expanding universe.
  
  The rest of the paper is organized as follows: in section ~\ref{sec:stdModel} some basic distance definitions of the \textit{standard model} are reviewed. The luminosity-angular distance for an expanding universe from the observer point of view is derived in section ~\ref{sec:ObserverPointOfView}. 
  
\begin{figure}
    \centering
    \leavevmode
      \includegraphics[width=0.5\textwidth]{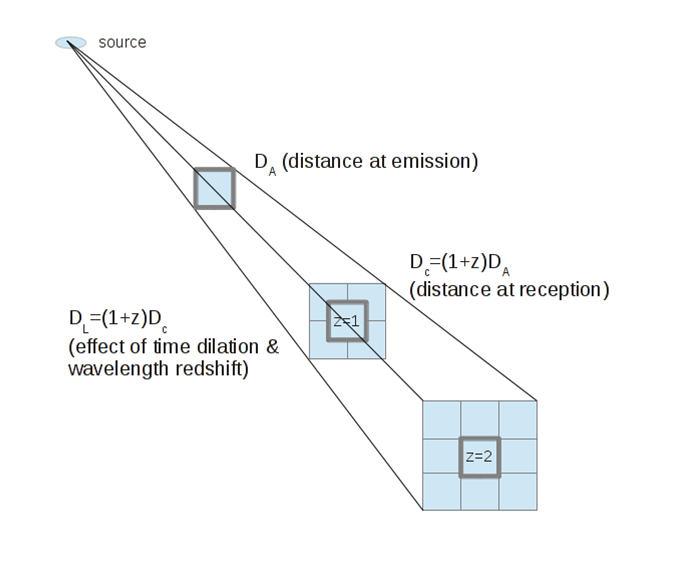}
      \caption{\textit{Standard view luminosity-angular distances relation}: \textit{Angular diameter distance} ($D_A$), \textit{comoving distance} ($D_C$) and \textit{luminosity distance} ($D_L$) for a flat universe. $D_A$ is the distance at emission, $D_C$ is the distance at reception and $D_L$ account for the distance elongation due to universe expansion ($\sim(1+z)$), time dilation and wavelength redshifting ($\sim(1+z))$. The relation $D_L=D_A(1+z)^2$ can be deduced from the figure.
       \label{fig:gauss3D_short}}
 \end{figure}%
  
 \section{Standard model luminosity-angular distances relation}
\label{sec:stdModel}

\subsection{Cosmological distances}
\label{sec:distancesstdModel}

\noindent
The \textit{angular diameter distance} $D_A$ is defined as the ratio between the object physical size $S$ and its angular size $\theta$ 
 
\begin{eqnarray}
\label{eq:angularDiameterDistanceTh}
  D_A=\frac{S}{\theta}
\end{eqnarray}
 
 On the other hand, the \textit{luminosity distance} defines the relation between the bolometric flux energy $f$ received at earth from an object, to its bolometric luminosity L by means of

\begin{eqnarray}
\label{eq:fluxEnergy}
 f= \frac{L}{4\pi D_L^2} 
\end{eqnarray}

being

\begin{eqnarray}
\label{eq:luminosity2_vs_angular2}
D_L^2= D_A^2(1+z)^4
\end{eqnarray}

\noindent
There are four (1+z) factors affecting to flux energy diminution (Fig.~\ref{fig:gauss3D_short}). Two come from the elongation of the initial distance $D_A$ by a factor of $(1+z)$ due to the assumed universe expansion. This elongation dilutes the luminosity by $D_A^2(1+z)^2$ according to the inverse square law. Another factor comes from the time dilation expected in an expanding universe that reduces the photon emission/reception rate by $(1+z)^{-1}$. The last factor comes from the cosmological wavelength redshift that decrease the energy of photons by $(1+z)^{-1}$. Therefore, the luminosity-angular distance relation for an expanding universe is given by

\begin{eqnarray}
\label{eq:stDistancesRelation}
D_L=D_A(1+z)^2    
\end{eqnarray}

\section{Standard luminosity-angular distance relation derived from the observer point of view}
\label{sec:ObserverPointOfView}
\subsection{Angular size conservation}

\begin{figure}
    \centering
    \leavevmode
      \includegraphics[width=0.6\textwidth]{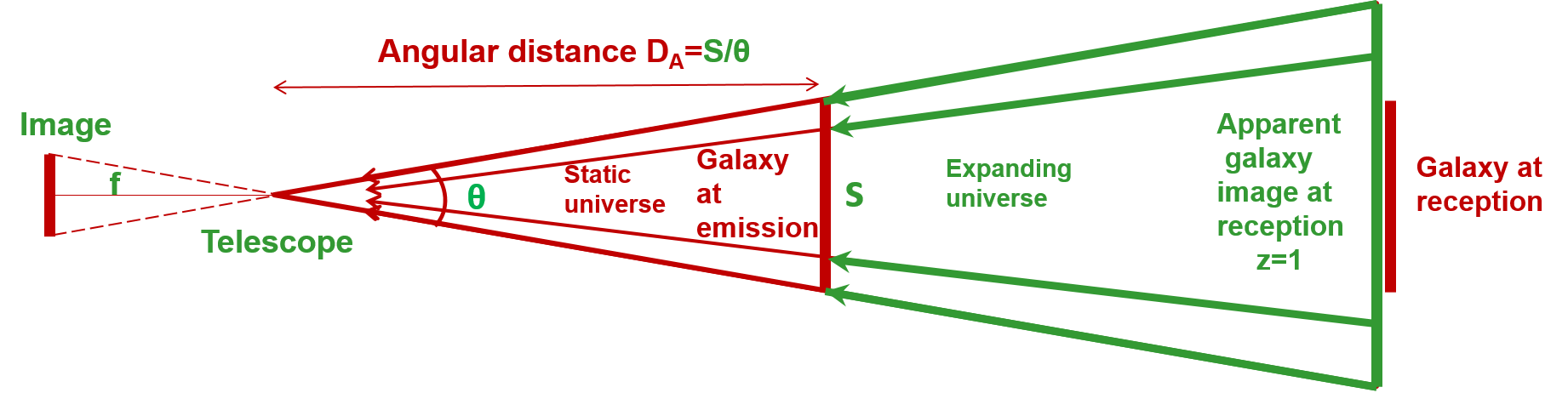}
      \caption{Size conservation: The apparent size of extra-galactic sources is the same in static and expanding universes. That agrees with the angular distance definition.}
       \label{fig:cl1_2D}
 \end{figure}%

\begin{figure}
    \centering
    \leavevmode
      \includegraphics[width=0.6\textwidth]{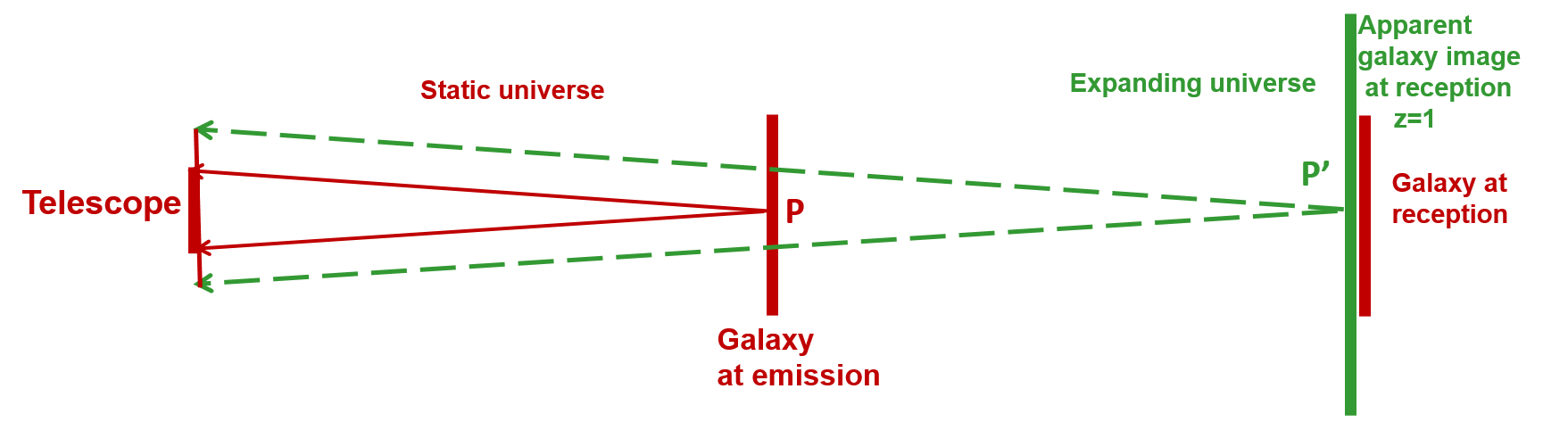}
      \caption{Flux dispersion: The light cone subtended from any point of the source would overflow the telescope surface in an expanding universe provoking a lost of flux-}
       \label{fig:fluxDispersion}
 \end{figure}%

Let us set the origin of coordinates at the observer O, and consider a extended cosmological object (galaxy) initially located at angular distance $D_A$ from O (Fig.~\ref{fig:cl1_2D}). The \textit{angular distance} is defined in cosmology as 

\begin{eqnarray}
  \label{eq:DA_definition}
   D_A=\frac{S}{\theta}
\end{eqnarray}

\noindent
where $S$ corresponds to the size of the object and $\theta$ the angle subtended by the object. Note that $D_A$ is defined identically for static and expanding universes. Since $D_A$ corresponds to the distance at emission its value is the same for both universes, so does $S$ and therefore $\theta$. Thus, since $\theta$ is the same for static and expanding universes, the objects are observed as they were at time of emission. The apparent image size (i.e., subtended angle) remains unaltered to the observer from the emission to reception in spite of the expansion. 

\subsection{Flux dispersion}
In previous version of this note, Fig.~\ref{fig:cl1_2D} was used to argument that the flux is focused towards the observer and it is not dispersed beyond $1/D_A^{2}$. Nevertheless, to assess for the received flux in an expanding universe we have to take into account the flux received not by a unique point (Fig.~\ref{fig:cl1_2D}) but by the telescope surface (Fig.~\ref{fig:fluxDispersion}). In Fig.~\ref{fig:fluxDispersion} we can see that the light cone spanning the telescope from any point of the source would be scaled due to the expansion, overflowing the telescope surface and hence dropping flux by $(1+z)^{-2}$ factor. This factor along with the one coming from time dilation and wavelength redshift sum up to $(1+z)^{-4}$. Therefore, the luminosity distance $D_L$ for an assumed expanding universe is 

\begin{eqnarray}
  \label{eq:fL_DL_DA_e}
   D_L=D_A(1+z)^2
\end{eqnarray}

while for a non-expanding universe would be 

\begin{eqnarray}
  \label{eq:fL_DL_DA}
   D_L=D_A(1+z)
\end{eqnarray}

whenever a mechanism is found for explaining the redshift and time dilation for this non.expanding case.

\bibliographystyle{unsrtnat}
\bibliography{el1}  

\end{document}